\documentclass[sigconf]{acmart}
\usepackage{booktabs} % For formal tables

\usepackage{multirow}
\usepackage{amsmath}
\usepackage{subfig}
\usepackage{enumitem}

\usepackage{multicol}
\usepackage{float}

\usepackage{hhline}
\usepackage{booktabs} % For formal tables
\usepackage{amssymb}
\usepackage{graphicx}
\usepackage{textcomp}
\usepackage[labelfont=bf]{caption}

\usepackage{algorithm}
\usepackage{algpseudocode}

\def\BibTeX{{\rm B\kern-.05em{\sc i\kern-.025em b}\kern-.08emT\kern-.1667em\lower.7ex\hbox{E}\kern-.125emX}}
    
\copyrightyear{2019}
\acmYear{2019} 
\acmConference[WWW '19 Companion]{Companion Proceedings of the 2019 World Wide Web Conference}{May 13--17, 2019}{San Francisco, CA, USA}
\begin{document}

%
% The "title" command has an optional parameter, allowing the author to define a "short title" to be used in page headers.
\title[Signed Link Prediction with Sparse Data]{Signed Link Prediction with Sparse Data: The Role of Personality Information}

%
% The "author" command and its associated commands are used to define the authors and their affiliations.
% Of note is the shared affiliation of the first two authors, and the "authornote" and "authornotemark" commands
% used to denote shared contribution to the research.

\author{Ghazaleh Beigi}
\affiliation{\institution{Computer Science and Engineering Arizona State University}}
\email{gbeigi@asu.edu}

\author{Suhas Ranganath}
\affiliation{\institution{Walmart Labs}}
\email{suhas.ranganath@walmartlabs.com}

\author{Huan Liu}
\affiliation{\institution{Computer Science and Engineering Arizona State University}}
\email{huan.liu@asu.edu}

%
% By default, the full list of authors will be used in the page headers. Often, this list is too long, and will overlap
% other information printed in the page headers. This command allows the author to define a more concise list
% of authors' names for this purpose.
\renewcommand{\shortauthors}{G. Beigi et al.}

%
% The abstract is a short summary of the work to be presented in the article.
\begin{abstract}
Predicting signed links in social networks often faces the problem of signed link data sparsity, i.e., only a small percentage of signed links are given. The problem is exacerbated when the number of negative links is much smaller than that of positive links. Boosting signed link prediction necessitates additional information to compensate for data sparsity. According to psychology theories, one rich source of such information is user's personality such as optimism and pessimism that can help determine her propensity in establishing positive and negative links. In this study, we investigate how personality information can be obtained, and if personality information can help alleviate the data sparsity problem for signed link prediction. We propose a novel signed link prediction model that enables empirical exploration of user personality via social media data. We evaluate our proposed model on two datasets of real-world signed link networks. The results demonstrate the complementary role of personality information in the signed link prediction problem. Experimental results also indicate the effectiveness of different levels of personality information for signed link data sparsity problem.
\end{abstract}



%
% Keywords. The author(s) should pick words that accurately describe the work being
% presented. Separate the keywords with commas.

\maketitle

\section{Introduction}
The rapid growth of social media has led to an increased amount of user generated data and accordingly the impediment of finding reliable information~\cite{beigi2018privacy,alvari2018early}. Positive links (e.g., trust and friendship relations) play an important role in helping online users find relevant and credible information~\cite{tang2013exploiting}. They have also been demonstrated to benefit many social media applications including recommendation and information filtering~\cite{xin2009social,beigi2018similar}. Similarly, negative links could help decision makers reduce uncertainty and vulnerability associated with decision consequences~\cite{cho2006mechanism,hardin2004distrust,larson2004distrust,mcknight2001trust}. Therefore, signed link prediction without negative links may result in a biased estimate of the effect of positive links~\cite{tang2014distrust}. Thus it is sensible to investigate both positive and negative links together in signed link prediction.

The problem of signed link prediction aims at inferring new positive/negative relations by leveraging existing ones. In recent years, the majority of the existing algorithms~\cite{chiang2011exploiting,leskovec2010predicting} use the topological structures and the properties of the existing signed links to make predictions. However, the available explicit positive links are often sparse and follow power-law distributions~\cite{tang2013exploiting}. The signed data sparsity problem gets worse as social media users tend to reveal more their positive disposition than their negative one; thus, negative links are often much sparser than positive links in a signed social network. To make a better signed link predictor, we need to overcome the data sparsity problem of the signed links.

As suggested by psychologists~\cite{asendorpf1998personality,burt1998personality}, user's optimism and pessimism are important factors that determine her propensity in establishing the positive/negative social relations. According to Scheier et. al.~\cite{scheier1985optimism}, a person is defined as {\it optimist} when she is more likely to reinterpret negative events in a positive way and find meaning and growth in stressful situations. On the other hand, an individual is referred to as {\it pessimist} when she is pre-occupied only with the negative aspects of the environment and overlooks the positive aspects~\cite{scheier1985optimism}.
%a more pessimistic individual might be preoccupied only with the negative or threatening aspects of the environment and overlook the positive or encouraging aspects
%Following the work of ~.... \cite{hecht2013neural,geers1998optimism}, a person is defined as optimist when she is more likely to give positive feedbacks to the given entity with the lower quality than the average in the current situation. While, an individual is defined as pessimist when she tends to give more negative feedbacks to the entities with the quality higher than average facing the current situation.
For example, optimistic users have better social functioning and relations. Therefore, they actively pursue social relationships and have higher chances in establishing positive links resulting in longer lasting friendships~\cite{geers1998optimism,scheier1985optimism}. In contrast, pessimists likely practice in the opposite way, i.e., having negative attitudes and expecting the worst of people and situations. Consequently, they often establish negative links with others~\cite{geers1998optimism,scheier2001optimism}. %Another interesting observation of optimism and pessimism is that people in general like optimists more than pessimists and thus react more positively to optimists than to pessimists. In other words, optimists likely attract more positive links while pessimists receive more negative links~\cite{forgeard2012seeing,carver1994effects,raikkonen1999effects,helweg2002stigma,brissette2002role}.
% \begin{figure*}[t]
% 	\centering
% 	\subfloat[Postive link prediction]{\includegraphics[scale=0.45]{fig3.PNG}}\quad
% 	\subfloat[Sign prediction]{\includegraphics[scale=0.45]{fig1.PNG}}\quad
% 	\subfloat[Signed link prediction]{\includegraphics[scale=0.45]{fig2.PNG}}
% 	\caption{\textbf{ An illustration of the differences of positive link prediction, sign prediction and signed link prediction}}\label{diff}
% \end{figure*}
People's personal traits can be observed on social media and serve as a good indicator of their personality~\cite{golbeck2011predicting}. This is because (1) social media websites allow for free interaction and open exchange of viewpoints, and (2) social media data can be aggregated to establish normative behavior of individuals. Previous research~\cite{beigi2016signed} has shown the correlation between users' personality information and positive/negative links in social networks. In particular it shows that:% and notices the following: % This work first exploited users' posted feedback on social media to estimate their optimism and pessimism. Next, it studied the relationship between users' optimism and pessimism scores and their tendency to establish and receive positive and negative links in social media networks. In particular, it shows that:
\begin{itemize}[leftmargin=*]
	\item Users with high optimistic behavior are more likely to establish and receive positive links than those with low optimism. %More optimistic users also have higher chances to receive positive links from other people than those with low optimism.
	\item Users who are more pessimistic are more likely to establish and receive negative links than those with low level of pessimism. %People tend to establish more negative links with users with higher pessimism level than those with lower level of pessimism.
\end{itemize}
These findings are in line with psychology research~\cite{geers1998optimism,scheier2001optimism,scheier1985optimism} and also suggest that user's personality, i.e., optimism and pessimism, may have potentials to alleviate  the signed link data sparsity problem and improve the performance of signed link prediction. We use optimism and pessimism as a representative aspect of personality.

Although previous research~\cite{beigi2016signed} has taken the first steps to study the correlation between users' personality and positive/negative links in signed social networks, it is still unclear how such information could be modeled mathematically and incorporated for predicting the positive and negative links, whether the information could help solving signed link data sparsity problem and what the impact of personality information is on signed link prediction problem. In this paper, we study the problem of signed link prediction by exploiting personality information, in particular, users' optimism and pessimism. In particular, we investigate how to leverage such information for positive and negative link prediction, and then we propose a novel signed link prediction framework \textit{SLP}. Our main contributions are as follows:
\begin{enumerate}[leftmargin=*]
	\item Provide a principled way to model optimism and pessimism information mathematically; %according to our findings in~\cite{beigi2016signed};
	\item Propose the framework \textit{SLP} which deploys personality information for the signed link prediction problem; and%is a low-rank approximation method. This framework seeks a low-rank representation of users and deploys personality information for the signed link prediction problem.; and
	\item Evaluate extensively \textit{SLP} on two datasets of real-world signed link networks and explore the impact of personality information on signed link prediction.%We evaluate \textit{SLP} extensively on two datasets of real-world signed link networks. We compare our method with various signed link predictors to evaluate the efficiency of \textit{SLP}. We also explore the impact of personality information on signed link prediction in terms of different strength levels of optimism and pessimism.
\end{enumerate}

% The remainder of the paper is organized as follows. In Section 2, we formally define the problem of exploiting personality information for signed link prediction. Next, we discuss how we measure users' optimism and pessimism from their feedback information~\cite{beigi2016signed}. Then, we detail the proposed method in Section 4. Section 5 presents the datasets and experiments and observations. We review the related work in Section 6. Finally we conclude the paper with presenting future work.
\section{Problem Statement}

We shall first assume $\mathcal{U} = \{u_1,u_2,\ldots,u_n\}$ is the set of $n$ users. We use the matrix $\bf{G}\in\mathbb{R}^{n\times n}$ to denote user-user positive and negative links where $\bf{G}_{ij}=1$, $\bf{G}_{ij}=-1$ and $\bf{G}_{ij}=0$ represent a positive link, a negative link, and missing (a.k.a unknown) information from $u_i$ to $u_j$ respectively. We also measure users' optimism and pessimism personality information following previous research~\cite{beigi2016signed}--this is discussed in more details in the next section. We follow psychology literature and consider two \textit{separate} measures for each user's optimism and pessimism as they should be conceptualized as two independent dimensions~\cite{fischer1986optimism,mroczek1993construct,beigi2016signed,chang1998distinguishing}. This means that being optimistic does not necessarily mean not pessimistic. Let us now assume vectors ${\bf o} \in \mathbb{R}^{n \times 1}$ and ${\bf p} \in \mathbb{R}^{n \times 1}$ denote the users' optimism and pessimism respectively, where ${\bf o}_{i} \in [0,1]$ is the optimism strength of $u_i$ and ${\bf p}_{i} \in [0,1]$ is her pessimism strength. The higher optimism (pessimism) score is, the more optimistic (pessimistic) $u_i$ is.

The problem of signed link prediction by exploiting user's optimism and pessimism is then formally defined as follows:\\
{\it Given users' optimism and pessimism ${\bf o}$ and ${\bf p}$, and user-user existing positive and negative links matrix ${\bf G}$, we seek to learn a predictor $\mathit{f}$ to find the unknown positive/negative link information by inferring new user-user positive/negative link matrix $\widetilde{\mathbf{G}}$ as follows:}
\begin{align}
	f:\{{\bf G},{\bf o}, {\bf p}\} \rightarrow \{\widetilde{\mathbf{G}}\}
\end{align}
The problem of signed link prediction is different than its existing variants, positive link prediction~\cite{tang2013exploiting} and sign prediction~\cite{yang2012friend} and more challenging compared to them. In particular, we aim to predict the existence of a link between a pair of nodes and its sign. We also predict both positive and negative links simultaneously. Besides, the vast majority of the existing work for the problem of signed link prediction~\cite{leskovec2010predicting} leverage only the existing links between users. In our work, we seek to leverage additional resources such as user's personality to alleviate the data sparsity problem.

% We illustrate the unique characteristics of the singed link prediction along with those of the existing variations, in Figure~\ref{diff}: %We list the key nuances of the signed link prediction problem as follows.
% \begin{itemize}[leftmargin=*]
% 	\item Positive link prediction (Figure~\ref{diff}(a)), seeks to predict only positive links from the existing ones. In contrast, as illustrated in Figure~\ref{diff}(c), in signed link prediction, we aim to predict both positive and negative links simultaneously.
% 	\item Sign prediction (Figure~\ref{diff}(b)) only infers the signs of the existing links. While, in the signed link prediction (Figure~\ref{diff}(c)), we predict the existence of a link between a pair of nodes and its sign.
% 	\item The vast majority of the existing work for the problem of signed link prediction~\cite{leskovec2010predicting} aims at predicting positive and negative links by leveraging only the existing links between users. This could result in an inaccurate link formation due to the data sparsity problem. In our, we seek to leverage additional resources such as user's personality to alleviate the data sparsity problem.
% \end{itemize}

\section{Computation of Personality Information}

Personality information is not readily available on social media. Individuals usually do not label themselves as optimistic or pessimistic. A conventional way of obtaining personality information is to directly ask people whether they expect outcomes in their lives to be good or bad~\cite{scheier1992effects}, which is often seen to use psychological surveys designed for measuring an individual's optimism and pessimism (e.g., \cite{scheier1994distinguishing}). However, since social media data is large-scale, and mainly observational, it is impractical to ask each and every user for their personality information. The onus is, therefore, on us to find a sensible way to infer if a user is optimistic or pessimistic or neither. An indirect approach is to measure optimism and pessimism based on the idea that people's expectancies for the future stem from their interpretations of the past~\cite{peterson1984causal}. Thus, past experiences can reflect an individual's levels of optimism or pessimism. With social media data, the question is how to define a computational measure of optimism or pessimism. To summarize, individuals do not explicitly offer their personality information, it is infeasible to ask a large number of users about such information, but individual social media users do leave their traces online. We ask if we can aggregate individual user's data and automatically figure out if a user is optimistic or not. %This makes the computation of personality information really challenging. we need to formalize how to measure different levels of users' optimism and pessimism. Measuring optimism and pessimism is a challenging task.

Scheier et. al. ~\cite{scheier1985optimism} defines optimism as re-interpreting negative events in a positive way and pessimism as preoccupying with the negative aspects and overlooking positive events.
%\subsection{Computation of Optimism and Pessimism}
Following the psychology literature, user's feedback could be also used to estimate her optimism and pessimism as they are counterparts of each other~\cite{hu2013exploring,hu2014exploring}. % In general, a viewpoint of an optimist/pessimist is completely different from that of a normal user, to the
It is shown in ~\cite{hu2013exploring,hu2014exploring} that on social media websites, optimists are more willing to give more positive feedback while pessimists are more biased toward giving more negative feedback than usual. In previous research ~\cite{beigi2016signed}, this observation has been utilized to calculate users' optimism and pessimism by leveraging their feedback to different entities in social media (e.g. posts and items). In this work, we use two different scenarios to illustrate how  personality scores can be computed. The two scenarios differ in their user feedback: (1) ratings given by users to items, or (2) opinions expressed by users towards each other. %~\cite{beigi2016signed}.
\subsection{Scenario 1: Ratings as Feedback} Here, feedbacks are users' ratings given to the items. %we assume entities and feedback to be items and ratings given by users to items, respectively.
Let $\mathcal{I}$ be the set of $M$ items and assume $r_{ik}$ denotes the rating score given from $u_i$ to $I_k$ with $r_{ik} = 0$ indicating that $u_i$ has not rated $I_k$ yet.  Also, consider $\overline{r}_k$ as the average rating score of the $k$-th item rated by users. Following~\cite{beigi2016signed}, we treat ratings less than some predefined threshold $r_{th}$ as low and above it as high. We use $\mathcal{O}_L(i)$ to denote the set of items with low average rating scores rated by $u_i$:
\begin{align}
	\mathcal{ O}_L(i) = \{I_k \mid r_{ik}\neq 0 \wedge \overline{r}_k\leq r_{th}\} \nonumber
\end{align}
$\mathcal{O}_{HL}(i)$ is further used to denote the set of items which have received high ratings from $u_i$, and meanwhile have low average scores.  $\mathcal{O}_{HL}(i)$ can be formally defined as:
\begin{align}
	\mathcal{O}_{HL}(i) = \{I_k \mid I_k \in  \mathcal{O}_{L}(i) \wedge {r}_{ik} > r_{th}\} \nonumber
\end{align}
Intuitively, the more frequently user $u_i$ has rated above the average, the more optimistic she is. Therefore, optimism score for $u_i$ is defined as ${\bf o}_i = \frac{|\mathcal{O}_{HL}(i)|}{|\mathcal{O}_{L}(i)|}$  where $|.|$ is the size of the set~\cite{beigi2016signed}.

Similarly, we use $\mathcal{P}_{H}(i)$ to denote the set of items with high average rating scores and rated by $u_i$,
\begin{align}
	\mathcal{P}_{H}(i) = \{I_k \mid r_{ik} \neq 0 \wedge \overline{r}_k > r_{th}\} \nonumber
\end{align}
Let $\mathcal{P}_{LH}(i)$ denotes the subset of items from $\mathcal{P}_{H}(i)$, which are given low rates by $u_i$:
\begin{align}
	\mathcal{P}_{LH}(i) = \{I_k \mid I_k \in  \mathcal{P}_{H}(i) \wedge {r}_{ik} \leq r_{th}\} \nonumber
\end{align}
The pessimism score for $u_i$ is defined as: ${\bf p}_i = \frac{|\mathcal{P}_{LH}(i)|}{|\mathcal{P}_{H}(i)|}$~\cite{beigi2016signed}.

\subsection{Scenario 2: Opinions as Feedback}  Here, feedbacks are users' opinions they expressed towards each other and individuals' personality is defined based on their positive/negative opinions~\cite{beigi2016signed}. Following the work of~\cite{beigi2016signed}, we create user-user positive and negative opinion matrices ${\bf P}$ and ${\bf N}$ by computing the number of positive or negative opinions users express toward each other. Let $\overline{P}$ and $\overline{N}$ be the average of positive and negative opinions between all pairs of users, respectively. We also define $\overline{{\bf P}}_j$ and $\overline{{\bf N}}_j$ as the average of positive and negative opinions received by $u_j$. Further, we define $\mathcal{O}_{L}(i)$, as a set of users $u_j$ who have received positive emotions from $u_i$, but at the same time, have received more negative emotions than the average of negative opinions in the network, i.e. they are among the worst people in the network:
\begin{align}
	\mathcal{O}_{L}(i) = \{u_j \mid {\bf P}_{ij}\neq 0 \wedge \overline{{\bf N}}_j> \overline{N}\} \nonumber
\end{align}
$\mathcal{O}_{HL}(i)$ denotes the set of users $u_k$ who belong to $\mathcal{O}_{L}(i)$ and have received more positive emotions from $u_i$ than $\overline{{\bf P}}_k$,
\begin{align}
	\mathcal{O}_{HL}(i) = \{u_k \mid u_k \in  \mathcal{O}_{L}(i) \wedge {\bf P}_{ik} > \overline{{\bf P}}_k\} \nonumber
\end{align}
Intuitively, the more frequently $u_i$ has given positive emotions to the worst users in the network, the more optimistic she is~\cite{beigi2016signed}. Thus, the optimism score for $u_i$ is defined as as ${\bf o}_i = \frac{|\mathcal{O}_{HL}(i)|}{|\mathcal{O}_{L}(i)|}$.

Likewise, we define $\mathcal{P}_{H}(i)$, as a set of users $u_j$ who have received negative emotions from $u_i$, but at the same time, have received more positive emotions than the average in the network, i.e. they are better than the average,
\begin{align}
	\mathcal{P}_{H}(i) = \{u_j \mid {\bf N}_{ij} \neq 0 \wedge \overline{{\bf P}}_j > \overline{P}\} \nonumber
\end{align}
We define $\mathcal{P}_{LH}(i)$ to denote the set of users $u_k$ who belong to $\mathcal{P}_{H}(i)$ and have received more negative emotions from $u_i$ than $\overline{{\bf N}}_k$,
\begin{align}
	\mathcal{P}_{LH}(i) = \{u_k \mid u_k \in  \mathcal{P}_{H}(i) \wedge {{\bf N}}_{ik} > \overline{N}_k\} \nonumber
\end{align}
Pessimism of $u_i$ could be similarly defined as: ${\bf p}_i = \frac{|\mathcal{P}_{LH}(i)|}{|\mathcal{P}_{H}(i)|}$~\cite{beigi2016signed}.
%Note also, there might be other ways to construct optimism and pessimism vectors, ${\bf o}$ and ${\bf p}$ such as incorporating text information.%psychological surveys which is beyond the scope of this paper.

\section{The Proposed Framework - SLP}

Readily available information such as user's personality and its impact on the formation of signed links, motivates us to exploit it to overcome the inherent sparsity of the signed links. We model this information mathematically and then incorporate it for predicting positive and negative links. Here, we introduce our approach for modeling user's optimism and pessimism and then detail the proposed method \textit{SLP} for signed link prediction.

\subsection{Basic Model for Signed Link Prediction}
Users usually establish signed links with only a few set of other users. This results in very sparse and low-rank networks. Therefore, low-rank approximation methods could be deployed for modeling signed links~\cite{hsieh2012low}. Moreover, the work of~\cite{hsieh2012low} showed that weak structural balance in signed networks leads to a low-rank approximation method for modeling the network. Let ${\bf U} = [{\bf U}_1, \ldots, {\bf U}_n]^\top \in \mathbb{R}^{n \times d}$ be the low-rank latent representations of users in $\mathcal{U}$ where ${\bf U}_i \in \mathbb{R}^d,~d \ll n$ is the low-rank latent vector representation of $u_i$. The matrix factorization model seeks a low-rank representation of ${\bf U}$ via solving the following optimization problem:
\begin{equation}
	\min_{\mathbf{U,V}}~~ \|\mathbf{W}\odot(\mathbf{G}-\mathbf{UVU}^\top)\|_\mathbf{F}^2+\lambda_{1}||\mathbf{U}||_\mathbf{F}^2+\lambda_{2}||\mathbf{V}||_\mathbf{F}^2,
\end{equation}
\noindent where $||\:.\:||_F$ is the Frobenius norm of a matrix, $\odot$ is Hadamard product, and $\mathbf{V}\in\mathbb{R}^{d\times d}$ captures the correlations among user latent representations. ${\bf W}_{ij}$ controls the contribution of ${\bf G}_{ij}$ in the learning process. A typical choice of ${\bf W}$ is to set ${\bf W}_{ij}=1$ if ${\bf G}_{ij}\neq0$, and ${\bf W}_{ij}=0$ otherwise. Two smoothness regularization terms are also added to avoid over-fitting where $\lambda_{1}$ and $\lambda_{2}$ are non-negative regularization parameters on $\mathbf{U}$ and $\mathbf{V}$, respectively. This standard model is very flexible to add prior knowledge from side information. In the following subsection,  we introduce how to incorporate user personality information into the standard model.
\subsection{Modeling User Personality Information}
%The analysis in Section~\ref{analysis} suggests the correlation between users' personality and positive and negative links.
%The component to capture user's personality is based on the
The analysis in the previous research~\cite{beigi2016signed} introduces two important findings according to users' optimism and pessimism: (a) users who are more optimistic tend to create and receive more positive links in comparison to the users with low level of optimism; and (b) users with more pessimism personality are more likely to create and receive more negative links compared to the users who are less pessimistic. To model the user's optimism and pessimism effect, we consider the following two cases for each pair of users $\langle u_i,u_j\rangle$:
\begin{itemize}[leftmargin=*]
	\item {\it Case 1:} $\mathbf{o}_i - \mathbf{o}_j >t_o$ and $d_i> d_j+\gamma_{ij}$;
	\item {\it Case 2:} ${\bf p}_j - {\bf p}_i > t_p$ and $d_j+\delta_{ij} < d_i$;
\end{itemize}
\noindent where $t_o$ and $t_p$ are thresholds to consider the significant optimism and pessimism difference between $u_i$ and $u_j$, respectively. We use $\gamma_{ij}$ and $\delta_{ij}$ to make the expected total degree comparisons flexible for $u_i$ and $u_j$.

In {\it Case 1}, $u_i$ has a higher level of optimism and $u_j$ has a lower level of optimism, then ${\bf o}_i - {\bf o}_j > t_o$,  which suggests that $u_i$ is more likely to receive positive links in comparison to $u_j$. This results in a higher degree for $u_i$, i.e., $d_i> d_j+\gamma_{ij}$. To account for this, we impose a penalty if $d_j-d_i+\gamma_{ij}$ is lesser than 0, when ${\bf o}_i - {\bf o}_j >t_o$. Similarly, in {\it Case 2}, if $u_j$ has a higher pessimistic level while $u_i$ has a lower pessimism, then $\bf{p}_j - \bf{p}_i >t_p$. This suggests that $u_j$ is more likely to receive negative links in comparison to $u_i$ which results in a higher negative degree (and consequently a lower total degree) for $u_j$, i.e., $d_j+\delta_{ij} < d_i$. Therefore, we impose a penalty if $d_j-d_i+ \delta_{ij}$ is greater than 0, when ${\bf p}_j - {\bf p}_i >t_p$. These two cases align well with the findings in~\cite{beigi2016signed} regarding optimism and pessimism of users. Therefore, we propose the following personality regularizations:
\begin{align}\label{RegTerm1}
	\sum_{\substack{(i,j), \\ {\bf o}_i -{\bf o}_j >t_o}} \max (0, d_j-d_i+\gamma _{ij}) ^2,
\end{align}
\begin{align}\label{RegTerm2}
	\sum_{\substack{(i,j), \\ {\bf p}_j -{\bf p}_i >t_p}} \max (0,  d_j-d_i+\delta_{ij}) ^2,
\end{align}
where $d_i = {\bf U}_i{\bf V}{\bf U}^\top\cdot{\bf 1}_{n\times 1}$. Next, we will show that by minimizing Eq.~\ref{RegTerm1} and Eq.~\ref{RegTerm2}, we can model personality information:% as the following,

\begin{itemize}[leftmargin=*]
	\item Eq.~\ref{RegTerm1} imposes a penalty when the user contradicts optimism behavior, where $d_j-d_i+{\gamma} _{ij} > 0$ and ${\bf o}_i -{\bf o}_j >t_o$.  Minimizing $d_j-d_i+{\bf \gamma} _{ij}$ will force $d_i$ to be closer to higher values and $d_j$ to be close to lower values. Thus, $u_i$ is more likely to establish positive links resulting in an increase in its degree while $u_j$ is less likely to have positive links so that its degree will be decreased.
	\item Eq.~\ref{RegTerm2} imposes a penalty when the user contradicts pessimism behavior, where $d_j-d_i+{\bf \delta} _{ij} > 0$ and ${\bf p}_j -{\bf p}_i >t_p$. Minimizing $d_j-d_i+{\delta} _{ij}$ will force $d_j$ to be closer to lower values and $d_i$ to be closer to higher values. Therefore, $u_j$ is more likely to establish negative links toward others resulting in a decrease in its degree while $u_i$ is less likely and as a consequent its degree will increase.
\end{itemize}
The above observations show that we can model personality information by minimizing Eq.\ref{RegTerm1} and Eq.\ref{RegTerm2}.

\subsection{Personality Information for Signed Link Prediction}
Having added the personality regularizations, our proposed framework \textit{SLP} now seeks to solve the following optimization problem:

\allowdisplaybreaks
\begin{align}
	\label{formula:Opt1}
	&\min_{\textbf{U,V}} ||\textbf{W}\odot(\textbf{G}-\textbf{UVU}^\top)||_\textbf{F}^2+ \lambda_{1}||\textbf{U}||_\textbf{F}^2+\lambda_{2}||\textbf{V}||_\textbf{F}^2+ \\
	&\alpha\sum_{\substack{(i,j), \\ {\bf o}_i -{\bf o}_j >t_o}} \max (0, d_j-d_i+{\gamma}_{ij}) ^2+\beta\sum_{\substack{(i,j), \\ {\bf p}_j -{\bf p}_i >t_p}} \max (0, d_j-d_i+{\delta} _{ij}) ^2 \nonumber
\end{align}
where $d_i = {\bf U}_i{\bf V}{\bf U}^\top\cdot{\bf 1}_{n\times 1}$ and ${\bf 1}_{n\times 1}\in \mathbb{R}^{n \times 1}$ is a vector with all elements equal to 1. The additional information related to the optimism and pessimism of users alleviates the data sparsity problem for signed link prediction. If $u_i$ does not have any positive/negative links, learning her latent factor is impossible. However, if we have information on how much $u_i$ is optimistic or pessimistic, we still can learn ${\bf U}_i$ for $u_i$ via personality regularization.

Since the optimization problem in Eq.\ref{formula:Opt1} is not jointly convex with respect to $\mathbf{U}$ and $\mathbf{V}$, there is no neat closed solution for it due to existence of the $\max$ function. Therefore, it can be rewritten into its matrix form as:
\begin{align}
	\label{formula:Opt}
	&\min_{\textbf{U,V}} ||\textbf{W}\odot(\textbf{G}-\textbf{UVU}^\top)||_F^2+ \lambda_{1}||\textbf{U}||_\textbf{F}^2+\lambda_{2}||\textbf{V}||_F^2 \nonumber \\
	&+ \alpha \| \textbf{R}^{\gamma,k} \odot ( \textbf{D}-\textbf{D}^T+ \gamma) \|_F^2+ \beta \| \textbf{R}^{\delta,k} \odot (  \textbf{D}-\textbf{D}^T + {\bf \delta}) \|_F^2 
\end{align}
where $\mathbf{D}={\bf 1}_{n \times n}\cdot{\bf U} {\bf V}^\top {\bf U}^\top$. We define ${\bf R}^{\alpha,k} \in \mathbb{R}^{n \times n}$ and ${\bf R}^{\beta,k} \in \mathbb{R}^{n \times n}$ in the $k$-th iteration,
\begin{align}
	{\bf R}^{\gamma,k}_{ij}=&\begin{cases}
		1,& \text{if }  {\bf o}_i -{\bf o}_j >t_o \text{ and } d_j+{\bf \gamma} _{ij} > d_i\\
		0,              & \text{otherwise}
	\end{cases},\nonumber \\
	{\bf R}^{\delta,k}_{ij}=&\begin{cases}
		1,& \text{if }  {\bf p}_j -{\bf p}_i >t_p \text{ and } d_j+\delta_{ij}>d_i\\
		0,              & \text{otherwise}
	\end{cases}.\label{R}
\end{align}
Expanding the objective function of Eq.~\ref{formula:Opt} in the $k$-th iteration can be written as,
\begin{equation}\label{formula:Opt2}
\min_{\textbf{U},\textbf{V}}\mathcal{J}_k =Tr(\mathbf{W}\odot(-2\mathbf{G}^\top+\mathbf{ U}\mathbf{ V}^\top \mathbf{U}^\top) \mathbf{UV}\mathbf{U}^\top) +\alpha \textit{\textbf{h}}(\gamma)+\beta \textit{\textbf{h}}(\delta)
\end{equation}
\noindent where the function $\textit{\textbf{h}}(x)$ is defined as,
\begin{align}
	&\textit{\textbf{h}}(x) =Tr(\textbf{R}^{x,k} \odot \big((\textbf{D}-\textbf{D}^T+x)(\textbf{D}^T+x^\top) +(\textbf{D}^T-\textbf{D}-x)\textbf{D}\big)),
\end{align}
where $x$ could be either $\gamma$ or $\delta$. We use the gradient descent method to solve Eq.~\ref{formula:Opt2}, which has been proven to gain an efficient solution~\cite{hsieh2012low}. The partial derivations of $\mathcal{J}_k$ w.r.t. $\mathbf{U}$ and $\mathbf{V}$ are,
\begin{align}\label{divU}
	\frac{1}{2}\frac{\partial {\mathcal{J}_k}}{\partial \mathbf{U}}=&\mathbf{W}\odot((\mathbf{UVU}^\top-\mathbf{G})\mathbf{UV}^\top)+\mathbf{W}\odot((\mathbf{UV}^\top\mathbf{U}^\top-\mathbf{G}^\top)\mathbf{UV})+\nonumber\\
	&\alpha(\mathbf{I}f(\gamma)\mathbf{UV}+f(\gamma)\mathbf{IUV}^\top)+\beta(\mathbf{I}f(\delta)\mathbf{UV}+f(\delta)\mathbf{IUV}^\top)\nonumber\\
	\frac{1}{2}\frac{\partial {\mathcal{J}_k}}{\partial \textbf{V}}=&\textbf{U}^\top( \mathbf{W}\odot(\textbf{UVU}^\top-\mathbf{G}))\textbf{U}-\alpha\mathbf{U}^\top f(\gamma)\mathbf{IU}-\beta\mathbf{U}^\top f(\delta)\mathbf{IU}\nonumber\\
\end{align}
where
\begin{align}
	&\textit{\textbf{g}}(x) =({\bf R}^{x,k} \odot (\mathbf{D}+x))^\top+ {\bf R}^{x,k} \odot \mathbf{D}^\top,\nonumber \\
	&I={\bf 1}_{n \times n},~~~~~~~~~~~~~~~~~~~~~~~~~~~~~~~~~~~~~~~~~~~~~~~~\nonumber \\
	&f(x)=g(x)^\top-g(x) ~~~~~~~~~~~~~~~~~~~~~~~~
\end{align}

With the partial derivations of $\mathbf{U}$ and $\mathbf{V}$, an optimal solution of the objective function in Eq.~\ref{formula:Opt} can be obtained as shown in Algorithm~\ref{alg:alg1}.
\begin{algorithm}[t]
	\caption{\textbf{The Proposed Framework SLP}}\label{alg:alg1}
	
	\begin{algorithmic}[1]
		\Require $\mathbf{G}$, $\mathbf{o}$, $\mathbf{p}$, \{ ${\delta}$, ${\gamma}$, $\lambda_{1}$,$\lambda_{2}$,$\alpha$,$\beta$,$d$,$t_o$, $t_p$\}.
		\Ensure $\widetilde{\mathbf{G}}$
		\State \text{Initialize $\mathbf{U}$ and $\mathbf{V}$ randomly}
		\While{\text{Not convergent}}
		\State \text{Calculate $\mathbf{R}^{\gamma,k}_{ij}$ and $\mathbf{R}^{\delta,k}_{ij}$ according to Eq.~\ref{R}}
		
		\State \text{Calculate $\frac{\partial {\mathcal{J}_k}}{\partial \mathbf{U}}$ and $\frac{\partial {\mathcal{J}_k}}{\partial \mathbf{V}}$}\text{accoording to Eq. \ref{divU}}
		
		\State \text{Update} $\mathbf{U}\leftarrow \mathbf{U}-\gamma_u\frac{\partial {\mathcal{J}_k}}{\partial \mathbf{U}}$
		
		\State \text{Update} $\mathbf{V}\leftarrow \mathbf{V}-\gamma_v\frac{\partial {\mathcal{J}_k}}{\partial \mathbf{V}}$
		\EndWhile
		\State \text{Set $\widetilde{\textbf{G}} = \textbf{UVU}^\top$}
	\end{algorithmic}
\end{algorithm}
The inputs to this algorithm are user-user positive/negative link matrix $\mathbf{G}$, expected degree difference matrices ${\bf\delta}_\alpha$ and ${\bf\delta}_\beta$ and users' optimism $\mathbf{o}$ and pessimism $\mathbf{p}$ scores. We randomly initialize $\mathbf{U}$ and $\mathbf{V}$ in line 1. From line 2 to 8, we update $\mathbf{U}$ and $\mathbf{V}$ until they converge. The algorithm will stop when objective function has little change or when it reaches a predefined maximal iterations. The output of algorithm is the estimated positive/negative link matrix which is computed as $\widetilde{\mathbf{G}} = \mathbf{UVU}^\top$. $|\widetilde{\mathbf{G}_{ij}}|$ shows the likelihood of establishing relation from $u_i$ to $u_j$ and $sign(\widetilde{\mathbf{G}_{ij}})$ also shows whether the relation is a positive or negative link.% where $sign(x) =1$ for $x > 0$, $sign(x) = -1$ for $x < 0$ and $ sign(x) = 0$ if $x = 0$.
%%%TODO\vspace{-5pt}
\subsection{Time Complexity}
The most time-consuming operations are the calculations of $\frac{\partial {\mathcal{J}}}{\partial \mathbf{U}}$ and $\frac{\partial {\mathcal{J}}}{\partial \mathbf{V}}$. We only discuss the time complexity analysis of these two steps. Let the number of non-zero elements of $\mathbf{W}$ be $N_w$. %The time complexities of computing $\frac{\partial \mathcal{J}}{\partial \mathbf{U}}$ and $\frac{\partial \mathcal{J}}{\partial \mathbf{V}}$ are given below:

\begin{itemize}[leftmargin=*]
	\item We first focus on the analysis of ${\bf R}^{\gamma,k}$ which needs to be calculated in each iteration $k$ and is $\mathcal{O}(N_und+nd^2+{N_t}^2)$. We can keep the difference between users' optimism scores- $|{\bf o}_i - {\bf o}_j|$ before running the algorithm.  If $N_t$ is considered the number of pair of users $|{\bf o}_i - {\bf o}_j|>$$t$, we only need to check whether $d_j+{\gamma} _{ij} > d_i$ is satisfied for these pairs or not. Due to the sparsity of ${\bf U}_i{\bf V}{\bf U}^\top$, the time complexity of calculating $d_i = {\bf U}_i{\bf V}{\bf U}^\top\cdot{\bf 1}_{n\times 1}$ is $\mathcal{O}(N_und+nd^2)$ if $N_u$ is the number of non-zero elements of ${\bf U}_i{\bf V}{\bf U}^\top$. The time complexity for comparing $N_t$ number of users is also ${N_t}^2$. The same analysis can be done for ${\bf R}^{\delta,k}$. Therefore, the total time complexity of calculating ${\bf R}^{\gamma,k}$ and ${\bf R}^{\delta,k}$ in each iteration is $\mathcal{O}(N_und+nd^2+{N_t}^2)$.
	\item For the analysis of $\frac{\partial \mathcal{J}}{\partial \mathbf{U}}$ in Eq.\ref{divU}, since $\mathbf{W}$ is very sparse, the time complexity of calculating both $\mathbf{W}\odot((\mathbf{UVU}^\top-\mathbf{G})\mathbf{UV}^\top)$ and $\mathbf{W}\odot((\mathbf{UV}^\top\mathbf{U}^\top-\mathbf{G}^\top)\mathbf{UV})$ are $\mathcal{O}(nd^2 + N_wd)$ in which $N_w$ is the number of non-zero elements of $\mathbf{W}$. Similarly, since ${\bf D}+{\bf \gamma}$ and ${\bf D}+{\bf \delta}$ only need to be computed once, we can get the time complexity of $\mathbf{I}f(\gamma)\mathbf{UV}+f(\gamma)\mathbf{IUV}^\top$ and $\mathbf{I}f(\delta)\mathbf{UV}+f(\delta)\mathbf{IUV}^\top$ as $\mathcal{O}(N_und+N_{\gamma,k}n + nd^2 )$ and $\mathcal{O}(N_und+N_{\delta,k}n + nd^2 )$, respectively. $N_{\gamma,k}$ and $N_{\delta,k}$ denote the numbers of non-zero elements of ${\bf R}^{\gamma,k}$ and ${\bf R}^{\delta,k}$, respectively. For brevity, we omit the detailed analysis of these terms. The time complexity of $\frac{\partial \mathcal{J}}{\partial \mathbf{U}}$ in each iteration is thus $\mathcal{O}(nd^2 + N_und+ (N_w + N_{\delta, k} + N_{\gamma, k})n)$.
	\item Now we discuss the time complexity of $\frac{\partial \mathcal{J}}{\partial \mathbf{V}}$ in Eq.\ref{divU}. Since the number of non-zero elements of $\mathbf{W}$ is $N_w$, the cost of calculating $\mathbf{W}\odot(\mathbf{UVU}^\top-\mathbf{G})$ is $\mathcal{O}(nd^2 +N_wd)$. Thus, the time complexity of $\mathbf{U}^\top( \mathbf{W}\odot(\mathbf{UVU}^\top-\mathbf{G}))\mathbf{U}$ is $\mathcal{O}(nd^2+N_wd)$. With the same approach for calculating the time complexities of $\mathbf{I}f(\gamma)\mathbf{UV}+f(\gamma)\mathbf{IUV}^\top$ and $\mathbf{I}f(\delta)\mathbf{UV}+f(\delta)\mathbf{IUV}^\top$, the time complexity of computing $\mathbf{U}^\top f(\gamma)\mathbf{IU}$ and $\beta\mathbf{U}^\top f(\delta)\mathbf{IU}$ is $\mathcal{O}(N_und+N_{\gamma,k}n + N_{\delta,k}n+ nd^2 )$. Hence, the total time complexity of $\frac{\partial \mathcal{J}}{\partial \mathbf{V}}$ in each iteration is $\mathcal{O}(nd^2 + N_und+ (N_w + N_{\delta, k} + N_{\gamma, k})n)$.
\end{itemize}
%%%TODO\vspace{-10pt}
\section{Experiments}
In this section, we first introduce the datasets we use, and then explore the effectiveness of \textit{SLP} for signed link prediction. In particular, we design an extensive set of experiments to (1) evaluate the proposed method \textit{SLP}, (2) investigate the effect of personality information on the performance of \textit{SLP} and (3) conduct parameter analysis to examine the sensitivity of \textit{SLP} to the main parameters.
%%%TODO\vspace{-10pt}
\subsection{Datasets}
We collect two large datasets of online signed social networks, Epinions and Slashdot. We perform some standard preprocessing steps by filtering out users without either positive or negative links. Table \ref{Tab:DataStat} shows the statistics of our datasets.

\textbf{Epinions.} this is a product-review website where users can establish trust and distrust relationships. We treat each trust and distrust relation as positive and negative links and construct user-user matrix ${\bf G}$ where ${\bf G}_{ij} = 1$ if user $i$ trusts user $j$, and ${\bf G}_{ij} = -1$ if user $i$ distrusts user $j$. Also, ${\bf G}_{ij} = 0$ if the information is missing. Users can rate each item in a scale of 1 to 5. Here we assume $r_{th}=3$, i.e., scores in $\{1,2,3\}$ are treated as low and $\{4,5\}$ as high scores~\cite{tang2014predictability}. We define optimism and pessimism based on the user's rating behavior in Epinions, following the scenario 1~\cite{beigi2016signed}.

\textbf{Slashdot.} this is a technology-related news platform which allows users to tag each other as either `friend' or `foe'. Similar to the Epinions, we construct user-user matrix ${\bf G}$ from the positive (friendship) and negative links (foes) in the network. Likewise, users can express their opinions toward each other by annotating the articles posted by each other. We define a user's optimism and pessimism based on her opinion expressing behavior following scenario 2~\cite{beigi2016signed}.
\begin{table}[t]%%TODO\vspace{-10pt}
	%\tbl{Statistics of the data \label{tab:data}}{
	\centering
		\small
		
	\caption{\textbf{Statistics of the datasets.}}\label{Tab:DataStat}
	%%TODO\vspace{-5pt}
	\begin{tabular}{l|l|l}
		& Epinions & Slashdot  \\ \hline \hline
		\# of Users & 131,828 & 7,275\\
		\# of Positive Links & 717,677 & 67,705 \\
		\# of Negative Links & 123,705 & 20,851 \\		
		\# of Ratings/Opinions & 577,692 & 403,896\\
		%	\# of Negative Opinions & 319,908 & 42,260\\ \hline
	\end{tabular}%%TODO\vspace{-10pt}
\end{table}
%%TODO\vspace{-5pt}
\subsection{Experimental Settings}
We use 5-fold cross validation for evaluation. Each time, we hold one fold out and treat it as our test set $\mathcal{A}$. From the remaining 4 folds, we pick $x\%$ of positive and $x\%$ of negative links to construct our training set $\mathcal{T}$. Then we set ${\bf G}_{ij} =0$, $\forall\langle u_i,u_j\rangle\in\mathcal{T}\cup\mathcal{A}$ and new representation of ${\bf G}$ is fed to each predicator. In this paper, we vary $x$ as $\{50,60,70,80,90,100\}$ to investigate how well our model performs with different sizes of training set.
%We randomly choose $x\%$ of trust and distrust relations for training as existing trust relations $\mathcal{O}$ and treat the remaining $100 - x\%$ of them as new trust and distrust relations $\mathcal{N}$ to be predicted. We remove trust and distrust relations in $\mathcal{N}$ by setting ${\bf G}_{ij} =0$, $\forall\langle u_i,u_j\rangle\in\mathcal{N}$ and the new representation of ${\bf G}$ is the input of each predicator. In this paper, we vary $x$ as $\{40,50,60,70,80,90\}$.
%Consider ${\bf T}=\{\langle u_i,u_j\rangle | {\bf G}_{ij} = 1)\}$ and ${\bf D}=\{\langle u_i,u_j \rangle | {\bf G}_{ij} = -1)\}$ as the set of users with trust and distrust relations, respectively.
%We use 5-fold cross validation. Each time, we choose $x\%$ of trust and distrust relations of 4 subsamples as training data. Then we treat the remaining fold for validation as test set by setting ${\bf F}_{ij} =0$, $\forall\langle u_i,u_j\rangle\in\mathcal{S}$ and the new representation of ${\bf F}$ is the input of each predicator. In this paper, we vary $x$ as $\{40,50,60,70,80,100\}$.
As stated before, in signed networks, positive links are often much denser than negative ones; hence signed links are imbalanced in both training and test sets. Therefore we rather use the Area Under Curve (AUC), to assess the performance of signed link predictors on predicted values of links between pairs of users in test set, $\{\widetilde{\mathbf{G}}_{ij} , \forall\langle u_i,u_j\rangle\in\mathcal{A}\}$.

To evaluate the proposed framework \textit{SLP}, we compare it with the following representative signed link predictors:
\begin{itemize}[leftmargin=*]
	\item \textbf{MF}~\cite{hsieh2012low}: This is a variant of the proposed method and preforms matrix factorization on ${\bf G}$. This baseline ignores the personality regularization. We select this method to see how signed link prediction method performs in absence of personality information.
	\item \textbf{DB/OP/RP}~\cite{shahriari2016sign}: This method first extracts two sets of topological-based features for each pair of users: the first set consists of seven (7) degree-based (DB) features, and the second set contains twelve (12) features describing user's optimism/reputation (OP/RP), which are derived from the links between users. A total of 19 features. Then, it trains a logistic regression classifier using these features to predict positive/negative links for a given pair.
	\item \textbf{All23}~\cite{leskovec2010predicting}: This method considers 23 different topological structure features for each pair of relation between users. It then trains a logistic regression classifier to predict positive/negative links. The deployed features could be categorized into two groups. The first category captures the local relations of a node to the rest of the network including positive/negative in-degrees and out-degrees. The second category is extracted according to the balance theory.
	%\item \textbf{All23+OP}: This method considers optimism and pessimism scores for each pair of users as four additional features added to the existing 23 features of {\it All23} in order to investigate how it performs by adding personal information features.
	\item\textbf{TDP}~\cite{guha2004propagation}: This baseline treats positive (trust)/negative (distrust) link propagation as a repeating sequence of atomic operations. In this propagation-based method, positive link propagates multiple steps while negative link propagates only a single step.
	\item \textbf{Random}: This algorithm assigns a random sign (+ or -) to each user pair.
	
\end{itemize}
%Although there are many trust predication methods available, they cannot be used for both trust and distrust prediction based on~\cite{tang2014distrust} where distrust is not negation of trust.
Note we do not compare \textit{SLP} with any traditional positive link predictor such as~\cite{jang2014trust,tang2013exploiting}. This is because the signed link prediction problem could not be carried out by trivially applying positive link predictors~\cite{tang2014distrust}, despite existence of several positive link predictors~\cite{jang2014trust,tang2013exploiting}. We use cross-validation to determine the best values for the proposed method and the baselines with parameters. For the proposed framework, we set the parameters as follows: $\{\lambda_{1}=\lambda_{2}=0.1, \:d=100, \:t_p=0.5, \:t_o=0.5, \:\alpha=80,\:\beta=80\}$. We also construct the matrices ${\bf \delta}^\alpha$ and ${\bf \delta}^\beta$ as follows,
\begin{align}
	{\bf \gamma}_{ij}=
	\begin{cases}
		15,& \text{if }  {\Delta\bf R^{\alpha}_{o_{ij}}} \geq 1500 \\
		10,& \text{if }  {\Delta\bf R^{\alpha}_{o_{ij}}} \geq 200 \\
		5,              & \text{otherwise}
	\end{cases}
	{\bf \delta}_{ij}=
	\begin{cases}
		15,& \text{if }  {\Delta\bf R^{\beta}_{p_{ij}}} \geq 500 \\
		10,& \text{if }  {\Delta\bf R^{\beta}_{p_{ij}}} \geq 150 \\
		5,              & \text{otherwise}
	\end{cases}
\end{align}
\noindent where ${\Delta\bf R^{\alpha}_{o_{ij}}}$ and ${\Delta\bf R^{\beta}_{p_{ij}}}$ indicate the difference between the ranks of users $u_i$ and $u_j$, assuming users are sorted in an descending order according to their optimism and pessimism scores, respectively.% Later in Section~\ref{param-analysis}, we will discuss the effect of changing the most important parameters, $\alpha$ and $\beta$, on the performance of SLP.
%More details about the impact of personality information on the performance of the proposed framework will be discussed in next subsections.
%Since the test set is selected randomly, the final results are reported by taking the average of 10 runs for each method.

\subsection{Performance Comparison}
The comparison results are shown in Table~\ref{Res_Ep} and Table~\ref{Res_Slash} for Epinions and Slashdot, respectively. We observe that all methods outperform the baseline Random. The performance of each method improves with the increasing size of the training data. \textit{SLP} performs the best among all methods. Next, we discuss why {\it SLP} did better:
\begin{itemize} [leftmargin=*]
	\item {\it SLP} outperforms {\it TDP} since edge signs could be incorporated into the signed link prediction rather than requiring a notion of propagation from farther-off parts of the network as~\cite{guha2004propagation} did. Moreover, in contrast to {\it SLP}, {\it TDP} does not consider imbalance distribution of positive/negative links.
	\item {\it SLP} always outperforms \textit{All23} since features extracted based on the topological structure, may not be robust due to the sparsity problem, and there might be many pairs of users without features based on balance theory~\cite{chiang2011exploiting}. The imbalance problem of positive/negative links distribution cannot be handled by \textit{All23}.
	\item {\it SLP} achieves better performance over {\it DB/OP/RP}, despite that both of the approaches leverage optimism/reputation-based features. The reason is {\it DB/OP/RP} uses topological structure to extract these features and hence suffers from the sparsity problem, similar to \textit{All23}. Simply put, there could be many pairs of users with zero optimism/reputation, which make leveraging the optimism/reputation-based features less useful in alleviating the imbalance problem of signed links distributions. In contrast, {\it SLP} infers users' personality information from their feedback on different issues other than merely using signed links.
	\item \textit{SLP} has better performance than {\it MF}. This is because {\it SLP} incorporates personality information to predict positive/negative links while {\it MF} does not. This suggests the importance of personality information in the problem of signed link prediction. %Though {\it All23+OP} leverages personality information, it cannot outperform {\it SLP} due to the larger number of structural features compared to the personality related ones.
\end{itemize}
We perform t-test on all comparisons and the t-test results suggest that all improvements are significant. To recap, the proposed framework obtains significant performance improvement by exploiting personality information.
\begin{table}[t]%%TODO\vspace{-10pt}
	\caption{\textbf{Performance comparison on Epinions.}}\label{Res_Ep}
	\centering	
	%	\includegraphics[width=0.5\textwidth]{Result3.png}
	%	\small
		%%TODO\vspace{-5pt}
	\begin{tabular}{p{1.2cm} p{0.8cm} p{0.8cm} p{0.8cm} p{0.8cm} p{0.8cm} p{0.8cm}}
		& \bf{50\%} & \bf{60\%} & \bf{70\%} & \bf{80\%} & \bf{90\%}  & \bf{100\%}\\ \hline
		\bf{SLP} & \textbf{0.8097} & \textbf{0.8113} & \textbf{0.8193} & \textbf{0.8258} & \textbf{0.8363} & \textbf{0.8504}\\ \hline
		\bf{MF} &  0.6497 & 0.6585 & 0.6614 & 0.6779 & 0.6884 & 0.6957\\ \hline
		\bf{DB/OP/RP} & 0.5981 & 0.6093& 0.6107 & 0.6202 &0.6379 &0.6426\\ \hline
		\bf{All23} & 0.5699 & 0.5721 & 0.5740 & 0.5797 & 0.5813 & 0.5934\\ \hline
		\bf{TDP} & 0.5461 & 0.5532 & 0.5591 &0.5613 & 0.5826  & 0.6001\\ \hline
		\bf{Random} & 0.4939 & 0.4994 & 0.5044 & 0.5024 & 0.5009 & 0.4997\\ \hline
	\end{tabular}%%TODO\vspace{-5pt}
	
\end{table}
\begin{table}[t]
	\caption{\textbf{Performance comparison on Slashdot.}}\label{Res_Slash}
	\centering	
	%	\includegraphics[width=0.5\textwidth]{Result3.png}
%		\small
		%%TODO\vspace{-5pt}
	\begin{tabular}{p{1.2cm} p{0.8cm} p{0.8cm} p{0.8cm} p{0.8cm} p{0.8cm} p{0.8cm}}
		& \bf{50\%} & \bf{60\%} & \bf{70\%} & \bf{80\%} & \bf{90\%} & \bf{100\%}\\ \hline
		\bf{SLP} &  \textbf{0.8308} & \textbf{0.8404} & \textbf{0.8464} & \textbf{0.8518} & \textbf{0.8613} & \textbf{0.8725}\\ \hline
		\bf{MF} &  0.6784 & 0.6810 & 0.6911 & 0.7066 & 0.7177 & 0.7265\\ \hline
		\bf{DB/OP/RP} & 0.6347 & 0.6401 & 0.6476 & 0.6558 	&0.6716 &0.6811\\ \hline
		\bf{All23} &  0.5913 & 0.5983 & 0.6005 & 0.6094 & 0.6126 & 0.6237\\ \hline
		\bf{TDP} &  0.5710 & 0.5794 & 0.5816 &0.5871 & 0.5981 & 0.6013\\ \hline
		\bf{Random} &  0.4924 & 0.5023 & 0.4981 & 0.4993 & 0.5003	& 0.5011\\ \hline
	\end{tabular}%%%TODO\vspace{-10pt}
	
\end{table}
\subsection{Further Experiment with Personality Information}
In this set of experiments, we probe further if the newly discovered personality-based features in {\it SLP} can be incorporated into other methods such as {\it All23} and {\it DB/OP/RP}. The addition of personality-based features to these two methods results two variants as follows.

\begin{itemize}[leftmargin=*]
	\item \textbf{All23+PI}~\cite{beigi2016signed,leskovec2010predicting}: This is a variant of {\it All23}~\cite{leskovec2010predicting} which considers four additional personality-based features (i.e., optimism and pessimism~\cite{beigi2016signed}) for each pair of users in addition to the existing 23 structural based features.
	\item \textbf{DB/OP/RP+PI}\cite{beigi2016signed,shahriari2016sign}: This is a variant of {\it DB/OP/RP}~\cite{shahriari2016sign} which uses the same four personality-based features~\cite{beigi2016signed} along with the existing topological-based features.
\end{itemize}

The comparison results are shown in Table~\ref{Res_Ep_2} and Table~\ref{Res_Slash_2} for Epinions and Slashdot. We observe the following:
\begin{itemize}[leftmargin=*]
	\item Both {\it All23+PI} and {\it DB/OP/RP+PI} outperform their corresponding original methods, i.e., {\it All23} and {\it DB/OP/RP}. This performance gain confirms the added value of the proposed personality features extracted from an exogenous source of information, in addition to those inferred from the signed links.
	\item Even with the personality-based features, {\it All23+PI} and \textit{DB/OP/RP+PI} were outperformed by {\it SLP} because {\it SLP} uses latent features for signed link prediction problem, while {\it All23+PI} and {\it DB/OP/RP+PI} both use a manually developed set of features.
\end{itemize}

%\item {\it All19} outperforms \textit{All23} because of considering new group of optimism/reputation based features. However, {\it All23+OP} beats {\it All19} which indicates that the proposed optimism/pessimism scores performs better than those of~\cite{shahriari2016sign}. The reasons are two fold, first the way~\cite{shahriari2016sign} defines optimism is not inline with psychological definitions~\cite{scheier1985optimism} and second~\cite{shahriari2016sign} extract optimism/reputation from topological structure and is not robust to the sparsity problem. Thus, there are many pairs of users with optimism/reputation scores of zero. Moreover, using topological features cannot help in handling imbalance problem of signed links distribution.
%\item \textit{SLP} always obtains the best performance. {\it SLP} incorporates personality information to predict positive/negative links while {\it MF} does not. This suggests the importance of personality information in the problem of signed link prediction. Though {\it All23+OP} leverages personality information, it cannot outperform {\it SLP} due to the larger number of structural features compared to the personality related ones.
\subsection{Impact of Personality Information on SLP}
We further explore the impact of personality information on signed link prediction, in an attempt to capture the different strength levels of optimism and pessimism that manifest online. First, we assume the size of training set is fixed to $100\%$ throughout this section. Then, using $k$-means, we divide the users in the training set into two groups, based on their optimism and pessimism scores: (1) $\mathcal{S}$: users with strong personality (both high optimism and pessimism), and (2) $\mathcal{I}$: indifferent users (with low optimism and pessimism).
\begin{table}[t]%%TODO\vspace{-10pt}
	\caption{\textbf{The impact of personality information on different signed link predictors on Epinions.}}\label{Res_Ep_2}
	\centering	
	%	\includegraphics[width=0.5\textwidth]{Result3.png}
	%	\small
		%%TODO\vspace{-5pt}
	\begin{tabular}{p{1.8cm} p{0.75cm} p{0.75cm} p{0.75cm} p{0.75cm} p{0.75cm} p{0.75cm}}
		& \bf{50\%} & \bf{60\%} & \bf{70\%} & \bf{80\%} & \bf{90\%}  & \bf{100\%}\\ \hline
		\bf{SLP} & \textbf{0.8097} & \textbf{0.8113} & \textbf{0.8193} & \textbf{0.8258} & \textbf{0.8363} & \textbf{0.8504}\\ \hline
		\bf{DB/OP/RP+PI} & 0.6761 & 0.6839& 0.6947 & 0.7014 &0.7177 &0.7263 \\ \hline
		\bf{DB/OP/RP} & 0.5981 & 0.6093& 0.6107 & 0.6202 &0.6379 &0.6426\\ \hline
		\bf{All23+PI} & 0.6531& 0.6673& 0.6701& 0.6848& 0.6974 & 0.7012\\ \hline	
		\bf{All23} & 0.5699 & 0.5721 & 0.5740 & 0.5797 & 0.5813 & 0.5934\\ \hline
		
	\end{tabular}%%TODO\vspace{-5pt}
	
\end{table}
\begin{table}[t]
	\caption{\textbf{The impact of personality information on different signed link predictors on Slashdot.}}\label{Res_Slash_2}
	\centering	
	%	\includegraphics[width=0.5\textwidth]{Result3.png}
	%	\small
		%%TODO\vspace{-5pt}
	\begin{tabular}{p{1.8cm} p{0.75cm} p{0.75cm} p{0.75cm} p{0.75cm} p{0.75cm} p{0.75cm}}
		& \bf{50\%} & \bf{60\%} & \bf{70\%} & \bf{80\%} & \bf{90\%} & \bf{100\%}\\ \hline
		\bf{SLP} &  \textbf{0.8308} & \textbf{0.8404} & \textbf{0.8464} & \textbf{0.8518} & \textbf{0.8613} & \textbf{0.8725} \\ \hline
		\bf{DB/OP/RP+PI} & 0.6938 & 0.7076 & 0.7188 & 0.7254 	&0.7374 &0.7513\\ \hline
		\bf{DB/OP/RP} & 0.6347 & 0.6401 & 0.6476 & 0.6558 	&0.6716 &0.6811\\ \hline
		\bf{All23+PI} &  0.6847& 0.6918& 0.7043& 0.7181& 0.7203 & 0.7328\\ \hline
		\bf{All23} &  0.5913 & 0.5983 & 0.6005 & 0.6094 & 0.6126 & 0.6237\\ \hline
	\end{tabular}
	
\end{table}

To assess the impact of personality information, we train our model on the whole training data, in three different ways by : (1) only considering the personality information (optimism and pessimism scores) of users in $\mathcal{S}$ (i.e., $\mathbf{o}_i=0,\mathbf{p}_i=0,  \forall u_i \in \mathcal{I}$), (2) only considering the personality of users in $\mathcal{I}$, (3) considering the personality of users in $\mathcal{S} \cup \mathcal{I}$. The results corresponding to these three runs are shown in Table~\ref{impact} with the following observations:
% We define the following instances of trust/distrust prediction problems as follows:
%\begin{itemize}
%	\item pTrust-O: we apply pTrust on the set of mostly optimistic users.
%	\item pTrust-P: in this instance, pTrust algorithm takes the set of mostly pessimistic users as an input.
%	\item pTrust-OP: here, users who has low optimistic and pessimistic scores are fed into the pTrust.
%\end{itemize}
%Without loss of generality and since these instances have almost the same size, we assume they are of exactly the same size.
\begin{itemize}[leftmargin=*]
	%	\item The removal of users with strong personality can negatively affect the performance of pTrust, comparing pTrust's accuracy on $\mathcal{I}$ (0.7486) and $\mathcal{S} \cup \mathcal{I}$ (0.8637); and
	%	\item The removal of indifferent users can positively help pTrust: $\mathcal{S}$ (0.9197) and $\mathcal{S} \cup \mathcal{I}$ (0.8637).	
	\item By removing users with strong personality, the performance of \textit{SLP} drastically drops. This can be observed by comparing the performance on $\mathcal{I}$ and $\mathcal{S} \cup \mathcal{I}$.
	\item The removal of indifferent users' personality does not have significant influence on the performance of \textit{SLP}, as we compare the performance on $\mathcal{S}$ and $\mathcal{S} \cup \mathcal{I}$.
\end{itemize}

%The role of personality information becomes clearer via this study: personality information can help trust and distrust prediction with a caveat: the stronger the personality, the more performance improvement. The caveat is that low personality information as indifferent users exhibit can lead to performance deterioration as shown in the performance on $\mathcal{I}$. In other words, pTrust suggests that it is difficulty to extract personality information from indifferent users  that can be relevant and useful in trust and distrust prediction.
\begin{table}[h]
	\caption{\textbf{The impact of personality information on SLP.}}\label{impact}
	\centering
	\subfloat[Epinions]{
	%\small
		\begin{tabular}{|c|c|c|} \hline
			$\bf{\mathcal{S}}$  & $\bf{\mathcal{I}}$ & $\bf{\mathcal{S} \cup \mathcal{I}}$ \\\hline
			{ 0.8189} & { 0.7677} & 0.8504\\\hline
		\end{tabular}}
		\hfill
		\subfloat[Slashdot]{
		%	\small
			\begin{tabular}{|c|c|c|} \hline
				$\bf{\mathcal{S}}$  & $\bf{\mathcal{I}}$ & $\bf{\mathcal{S} \cup \mathcal{I}}$ \\\hline
				{ 0.8406} & { 0.8094} & 0.8725\\\hline
			\end{tabular}}%%TODO\vspace{-10pt}
		\end{table}
The role of personality information becomes clearer via this study: personality information can help signed link prediction with a caveat-- the stronger the personality, the more improvement in performance. In other words, low personality information as indifferent users exhibit, can lead to performance deterioration as shown in the performance on $\mathcal{I}$. Moreover, with only a slight difference, the personality information from users in $\mathcal{S} \cup \mathcal{I}$, is worthless compared to the information from $\mathcal{S}$. This supports the fact that the information from users in $\mathcal{I}$ does not make much difference. In other words, the personality information from indifferent users could be treated as irrelevant to the signed link prediction problem.

		\subsection{Parameter Sensitivity Analysis of SLP}\label{param-analysis}
\begin{figure}[t]%%TODO\vspace{-10pt}
			\centering
			\subfloat[\bf{Epinions}]{\includegraphics[scale=0.40]{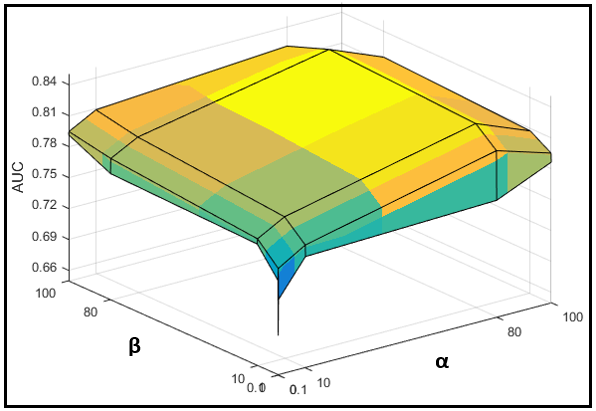}}\\%\quad%0.23
			\subfloat[\bf{Slashdot}]{\includegraphics[scale=0.40]{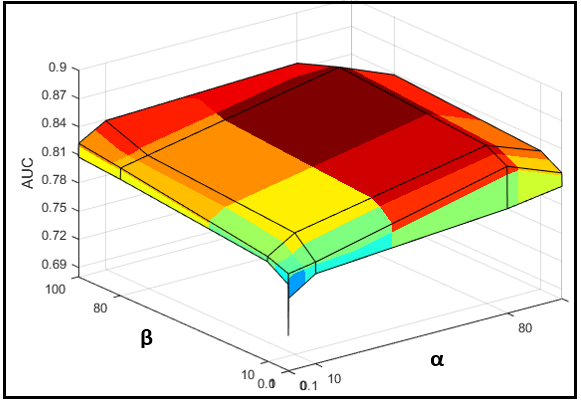}}
			\caption{\textbf{Effect of personality regularization in SLP.}}\label{fig:params22}%%TODO\vspace{-20pt}
		\end{figure}

		\textit{SLP} has two important parameters, $\alpha$ and $\beta$ which control the contribution from user's optimism and pessimism, respectively. Here, we discuss their effect by varying both $\alpha$ and $\beta$ as $\{0, 0.1, 10, 80, 100\}$ when the size of training set is fixed to $100\%$. The results are shown in Figure~\ref{fig:params22} with the followings observed.
		%\begin{figure*}[ht]
	
		\begin{itemize}[leftmargin=*]
			\item In general, the performance of \textit{SLP} increases with the increase of $\alpha$ and $\beta$,  and then it degrades. These patterns ease the parameter selection for \textit{SLP}.
			\item The performance improves even when $\alpha$ and $\beta$ slightly change from $0$ to $0.1$, which confirms the importance of user personality information in signed link prediction.
			\item After reaching to certain values, continuing to increase $\alpha$ and $\beta$ will result in the performance reduction. This suggests that large values of $\alpha$ and $\beta$ dominate the learning process and the model could learn $U$ and $V$ inaccurately due to the overfitting of the model to the personality information.
			\item \textit{SLP} is more sensitive to $\beta$ than $\alpha$ since $\alpha$ controls the contribution of user's optimism, which is denser than pessimism information.
		\end{itemize}
		%These results further demonstrate the importance of user's personality in the signed link prediction.
		\section{Related Work}
		The ease of using the Internet has raised numerous security and privacy issues. Mitigating these concerns has been studied from different aspects such as identifying malicious activities~\cite{alvari2019hawkes,alvari2018early,alvari2017semi}, addressing users' privacy issues~\cite{beigi2018securing,beigi2019protecting,beigi2018privacy} and finding positive/ negative links (i.e., trust/distrust) between users~\cite{tang2013exploiting,tang2015negative,beigi2016signed,beigi2016exploiting}. 
		Positive link prediction (a.k.a. trust prediction) has been extensively studied~\cite{jang2014trust,tang2013exploiting,beigi2014leveraging} in which the goal is to predict only positive links from existing ones. The availability of signed networks has motivated the research on signed link prediction~\cite{chiang2011exploiting,leskovec2010predicting}. The recent advances on signed link prediction demonstrate that negative links have added value in addition to positive links~\cite{tang2014distrust}. Thus, we focus on the positive/negative link prediction problem which is different from positive link prediction~\cite{tang2013exploiting} and sign prediction~\cite{yang2012friend,javari2014cluster}. It is more challenging compared to them because: first, in contrast to positive link prediction, in signed link prediction, we aim to predict both positive and negative links simultaneously and second, the sign prediction problem only infers the signs of the existing links while, in the signed link prediction, we predict the existence of a link between a pair of nodes as well as its sign.
		
		The existing signed link prediction studies can be divided into two categories: supervised and unsupervised methods. Supervised methods consider the signed link creation problem as a classification problem by using the old links and training a classifier on the features extracted from the signed networks\cite{chiang2011exploiting,leskovec2010predicting}. For example, the work of~\cite{leskovec2010predicting} first extracts in-degree and out-degree from signed links and then uses balance and status theory to extract triangle-based features and it then verifies the importance of balance and status theory for signed link prediction. Another work of~\cite{chiang2011exploiting} extends the triangle-based features to the k-cycle-based features. 
		
		Unsupervised methods often perform predictions based on certain topological properties of signed networks~\cite{hsieh2012low,symeonidis2013spectral,ye2013predicting}. One type of unsupervised methods is node similarity based methods~\cite{symeonidis2013spectral}, which first define similarity metrics to calculate node similarities, and then provide a way to predict the signed relations based on them. Another type of unsupervised methods is propagation-based methods~\cite{de2006many,guha2004propagation,ziegler2005propagation}. Positive sign propagation is treated as a repeating sequence of matrix operations~\cite{guha2004propagation}. Negative sign propagation then stopped after multiple steps of positive sign propagation~\cite{guha2004propagation}. The work of~\cite{de2006many} considers ignorance as well as partial positive/negative links in its proposed positive/negative link propagation method by modeling the network as an intuitive fuzzy relations. %Another work of~\cite{ziegler2005propagation} also proposes to integrate negative links into the process of the Appleseed positive links computation instead of superimposing it afterwards.
		Another category of unsupervised methods is based on low-rank matrix factorization~\cite{hsieh2012low,ye2013predicting,wang2017online}. The work in~\cite{hsieh2012low} models the signed link prediction problem as a low-rank matrix factorization model, based on the weak structural balance on the signed network. Another study in~\cite{ye2013predicting} extends the low-rank model to perform link prediction across multiple signed networks. Also the work of Wang et al. ~\cite{wang2017online} completes a binary matrix with positive and negative elements in an online setting by penalizing the difference between predicted matrix and the ground truth by logistic loss and matrix max-norm. Authors of ~\cite{tang2014predictability} incorporate side information, i.e., helpfulness ratings in a low-rank matrix factorization model to predict negative links. Beigi et al.~\cite{beigi2016exploiting} incorporates users' emotional information in a low-rank matrix factorization framework to predict positive and negative links between them. Another work of~\cite{tang2015negative} predicts the negative links using positive links and content-centric user interactions. 
		
		Likewise, \textit{SLP} is also based on the low-rank matrix factorization model. The difference between \textit{SLP} and the above low-rank matrix factorization models is that we investigate the role of personality information, i.e. optimism/pessimism in signed link prediction. Since optimists tend to establish more positive links than others and pessimists establish more negative links~\cite{geers1998optimism}, we model this fact as a constraint in the low-rank matrix factorization cost function to guide the learning process of $\mathbf{U}$ and $\mathbf{V}$.
		
		Exploiting user's features such as trustworthiness, bias, and optimism has been discussed in prior studies~\cite{shahriari2016sign,mishra2011finding}. For example, the work of~\cite{shahriari2016sign} addresses the problem of sign prediction based on users' optimism/reputation. The authors define optimism as users' voting pattern towards others and the reputation as their popularity. Specifically, their approach calculates the optimism as the difference between the number of user's positive and negative out-links. Moreover, they introduce rank based optimism and reputation based on the rank of users in the signed social network. Another work~\cite{mishra2011finding} computes bias and prestige of nodes based on the positive links between users in signed social networks. It defines bias as the user's truthfulness. The prestige is also calculated based on the opinion of other users in the form of inlinks a user gets.
		
		Similarly, we calculate users' optimism/pessimism and exploit these information for signed link prediction. The difference between our work and the above works is that we infer users' optimism/pessimism based on a source other than signed links, i.e. users' feedback and interactions on different entities (items and posts). These additional sources of information could help to overcome the data sparsity and imbalance problem for signed link data. %In particular, the sparsity problem will get even worse if we use users' link information to infer their optimism and pessimism.
		
		%Therefore exploiting optimistic/pessimistic role can mitigate the data sparsity problem and has potentials in improving the performance of signed link prediction.
		
		\section{Conclusion and Future Work}
		%We verified the impact of user's personality information on the formation of positive/negative links and then introduced personality regularization to capture those information mathematically. We then proposed the framework \textit{SLP} to infer positive/negative links by exploiting user's personality information. We evaluated \textit{SLP} on real-world datasets and the results demonstrated the complementary role of personality information in the signed link prediction problem.
		
		In this paper, we study the role of user personality, in particular, optimism and pessimism to mitigate the data sparsity problem in signed link prediction. User's feedback is further used to estimate optimism/pessimism. We then investigate the incorporation of such information for predicting positive/negative links and solving data sparsity problem for signed link prediction. We propose the framework \textit{SLP} by mathematically incorporating optimism and pessimism information. We evaluate \textit{SLP} on real-world datasets and the results demonstrate the complementary role of personality information in the signed link prediction problem. This work also enables us to experiment if incorporating different types of personality information can be relevant or helpful.
		
		One future direction is to consider additional user's personality traits including extraversion, agreeableness, openness, conscientiousness and neuroticism as suggested in the Big Five Model~\cite{mccrae1998introduction}. We also plan to explore more on inferring users personal information by incorporating content information. Furthermore, we can expand our statistic solution of user behavior to a dynamic one as users can evolve over time and they adapt to different situations, though slowly. A nuanced solution is to consider temporal dynamics of user's personality for dynamic signed link prediction. %Signed links have shown to benefit many applications such as recommendation systems. 
		 Another future direction is to explore the impact of users' personality and signed links on recommendation systems as signed links have previously shown promising results for recommendation task.

\begin{acks}
%	The authors would like to thank the anonymous referees for their valuable comments and helpful suggestions.
	 This material is based upon the work supported, in part, by NSF \#1614576, ARO W911NF-15-1-0328 and ONR N00014-17-1-2605.
\end{acks}
%
% The next two lines define the bibliography style to be used, and the bibliography file.
\bibliographystyle{ACM-Reference-Format}
%\bibliography{sample-base}

\end{document}